\documentclass[manuscript]{aastex}

\slugcomment{}
\shorttitle{Filament Eruption, Associated CME, and Coronal Downflows}
\shortauthors{Navin Chandra Joshi et al.}
\begin{document}
\title{Study of Failed CME Core Associated with Asymmetric Filament Eruption}
\author{Navin Chandra Joshi\altaffilmark{1}, Abhishek K. Srivastava\altaffilmark{1}}
	\affil{Aryabhatta Research Institute of Observational Sciences (ARIES), Manora Peak Nainital-263 002, India; navin@aries.res.in, njoshi98@gmail.com, aks@aries.res.in}
\author{Boris Filippov\altaffilmark{2}}
	\affil{Pushkov Institute of Terrestrial Magnetism, Ionosphere and Radio Wave Propagation, Russian Academy of Sciences, Troitsk, Moscow, Russia}
\author{Wahab Uddin\altaffilmark{1}}
	\affil{Aryabhatta Research Institute of Observational Sciences (ARIES), Manora Peak Nainital-263 002, India}
\author {Pradeep Kayshap\altaffilmark{1}}
	\affil{Aryabhatta Research Institute of Observational Sciences (ARIES), Manora Peak Nainital-263 002, India}
\author {Ramesh Chandra\altaffilmark{3}}
	\affil{Department of Physics, D.S.B. Campus, Kumaun University, Nainital-263 002, Uttarakhand, India}
\altaffiltext{1}{Aryabhatta Research Institute of Observational Sciences (ARIES), Manora Peak Nainital-263 002, India; navin@aries.res.in, njoshi98@gmail.com, aks@aries.res.in}
\altaffiltext{2}{Pushkov Institute of Terrestrial Magnetism, Ionosphere and Radio Wave Propagation, Russian Academy of Sciences, Troitsk, Moscow, Russia}
\altaffiltext{3}{Department of Physics, D.S.B. Campus, Kumaun University, Nainital-263 002, Uttarakhand, India}
\begin{abstract}
We present the multi-wavelength observations of asymmetric filament eruption, associated CME and coronal downflows on 2012 June 17-18 during 20:00-05:00 UT. We use SDO/AIA, STEREO-B/SECCHI observations to understand the filament eruption scenario and its kinematics. While LASCO C2 observations have been analyzed to study the kinematics of the CME and associated downflows. SDO/AIA limb observations show that the filament exhibits whipping like asymmetric eruption. STEREO/EUVI disk observations reveal a two ribbon flare underneath the south-eastern part of the filament that is most probably occurred due to reconnection process in the coronal magnetic field in the wake of the filament eruption. The whipping like filament eruption later gives a slow CME in which the leading edge and the core propagate respectively with the average speed of $\approx$ 540 km s$^{-1}$ and $\approx$ 126 km s$^{-1}$ as observed in the LASCO C2 coronagraph. The CME core formed by the eruptive flux-rope shows the outer coronal downflows with the average speed of $\approx$ 56 km s$^{-1}$ after reaching up to $\approx$4.33 $R_{\sun}$. Initially, the core decelerates with $\approx$ 48 m s$^{-2}$. The plasma first decelerates gradually up to the height of $\approx$4.33 $ R_{\sun}$ and then starts accelerating downward. We suggest a self-consistent model of a magnetic flux rope representing the magnetic structure of the CME core formed by eruptive filament that lost its previous stable equilibrium when reach at a critical height. With some reasonable parameters, and inherent physical conditions the model describes the non-radial ascending motion of the flux rope in the corona, its stopping at some height, and thereafter the downward motion, which are in good agreement with the observations.
\end{abstract}
\keywords{Sun: Corona - Sun: Filament - Sun: Magnetic field - Sun: Coronal Mass Ejection}
\section{Introduction}
\label{sec1}
Filaments/Prominences are the cool ($10^4$ K) and denser ($10^{10}-10^{11} cm^{-3}$) plasma material embedded
in the magnetic field of the ambient solar atmosphere (Mackay et al. 2010; Labrosse et al. 2010). Filaments and prominences 
are the same solar magnetic structures but appear distinctly due to their projection. Prominences, when observed on the disk, are known 
as solar filaments that lie along the polarity inversion line (Durrant 2002; Filippov \& Srivastava 2011). Filaments may remain in the quiescent state for 
several days over the solar disk, while some filaments may exhibit eruptive nature.
Filament eruption occurs mainly due to the non-equilibrium between the upward magnetic pressure and the downward 
magnetic tension, or due to the tether cutting between magnetic flux ropes containing filaments and overlaying 
arcades (Mackay et al. 2010). Filaments show different types of eruptions  such as failed eruption (Liu et al. 2009), partial eruption (Tripathi et al. 2009),  and full eruption (Schrijver et al. 2008; Chandra et al. 2010). Apart from these eruptions, filament can also undergo an asymmetric eruption with its one 
footpoint fixed in the photosphere (anchored leg) while the other participating in the dynamic motion (active leg) (Liu et al. 2009; Yang et al. 2012). 
Eruption of a flux rope can be partial or full that also depends on the overlying magnetic field configuration. 
Recently, Kumar et al. (2011) observed a huge flux-rope failed in the eruption, and found 
that the overlying filament remnant and coronal magnetic field caused its suppression. It is widely reported in the literatures 
that the emergence of new magnetic flux, generation of the high critical twist, low-lying atmospheric reconnections, can generate the 
eruptions and instability of the flux-ropes that may further lead to the outer coronal transients (Srivastava
et al. 2010; Chandra et al. 2011; Yan et al. 2012; Zhang et al. 2012; Botha et al. 2012;
Srivastava et al. 2013). 
However, it is also found that these conditions may only be the necessary conditions, which, may not be the sufficient conditions 
for the solar eruptions. The configuration of the ambient magnetic field as well as its strength may also play a role in deciding 
the nature and morphology of the eruptions (Kumar et al. 2011, reference cited
therein).

Eruptive filaments are closely associated with Coronal Mass Ejections (CMEs). CMEs are the propulsion of large-scale plasma 
and magnetic field into outer coronal and interplanetary space (Joshi et al. 2013). In general, a CME has a three part 
structure, i.e., a leading edge, dark cavity, and bright core associated with the eruptive filament 
(Riley et al. 2008, and reference therein). CMEs produce outward movement in the corona, and it may be either driven by the 
magnetic pressure or shock into the background of ambient solar wind (Filippov \& Srivastava 2010). 
The less denser outer corona sometime shows downflows of the plasma blobs, CME cores, etc. Such downflows of the magneto-plasma 
are collectively termed as "coronal inflows" or "coronal downflows". Various kinds of downflows/inflows into the solar corona 
have been observed and analyzed, which include the prominence fall back (Tripathi et al. 2006, 2007), dwonflow over the post flare 
arcades (Innes et al. 2003; McKenzie \& Hudson 1999; McKenzie 2000; Asai et al. 2004), inflow of small faint plasma structures and blobs, etc 
(Wang et al. 1999, 2000; Wang \& Sheeley 2002; Sheeley \& Wang 2002).

Innes et al. (2003) observed a series of dark, sunward moving flows, which were seen against the bright extreme ultraviolet 
post flare arcades of the large eruptive flare on 1999 April 21. Asai et al. (2004) examined the downflow motion above 
flare loops observed on 2002 July 23, and found that the plasma downflow occurred when the magnetic energy was 
released.  In all these examples, the reconnection is responsible for the downflows in the inner solar corona. In spite of such 
observed downflows in the inner corona, reconnection in the outer corona may also be responsible for downflows there. 
Wang et al. (1999) reported various small and fainter plasma flows moving through the corona, and found that these inward 
motions are the observational signature of the gradual closing-down of magnetic flux-tubes dragged outward by the CMEs 
or other transient outflows. Wang et al. (2000) studied the variety of small scale downflowing structures during the high solar 
activity (e.g., plasma blobs) and interpreted that these downflows were due to the magnetic reconnection between 
the closed and open field regions of the corona. Sheeley et al. (2001) reported that the inflow rate was dominated by transient 
bursts, which were correlated with the existence of non-polar coronal holes and other signatures of the solar non-axisymmetric 
open field structures. Wang \& Sheeley (2002) and Sheeley \& Wang (2002) identified faint, inward-moving features, e.g., collapsing loops, 
sinking plasma columns, falling plasma curtains, in and out pairs of oppositely directed plasma ejecta, and downflow of 
the core of CME at heliocentric distances from 2-6 $R_{\sun}$, and interpreted them as an initiation stage of the large-scale magnetic
reconnection. Tripathi et al. (2006, 2007) also reported coronal downflows in the course of prominence eruption and associated 
coronal mass ejection on 2000 March 5, and interpreted that the origin of such downflows may be due to the reconnection 
inside the bifurcating flux-rope.

As said above, the CME core some time exhibits downfall depending upon the local magnetic field and plasma 
configurations, and the most probable reason may be the pre-settings of the large-scale reconnection in the outer corona. 
The CME itself can also produce several changes in the outer coronal magnetic fields in form of the deflection of  
coronal streamers, kink propagation in coronal rays, etc (Filippov \& Srivastava 2010). Filippov \& Srivastava (2010)
 analyzed the events of the interaction of CMEs with the coronal rays observed by SoHO/LASCO, and interpreted these deflections as the 
influence of magnetic field of a moving flux rope associated with the CME on the remote coronal rays. In the present paper, 
we report multi-wavelength investigation of asymmetric filament evolution and eruption, associated CME, 
coronal downflows, and their relationship using the observations from Solar Dynamic Observatory/Atmospheric Imaging 
Assembly (SDO/AIA), Solar Terrestrial Relation Observatory/Sun Earth Connection Coronal and Heliospheric 
Investigation (STEREO/SECCHI) and Solar and Heliospheric Observatory/Large Angle and Spectrometer Coronagraph
(SoHO/LASCO) C2 instruments on 2012 June 17-18 during 20:00--05:00 UT. The observational results are presented in 
Section 2. Physical scenarios of the observations are discussed in Section 3. 
Discussion and conclusions are outlined in the last section.
\section{Observational Results}
\label{sec2}
We use multi-wavelength and multi-instrument data from SDO/AIA (Lemen et al. 2012), STEREO-B/SECCHI (Wuelser et al. 2004)
, and SoHO/LASCO C2 coronagraph (Brueckner et al. 1995) for this study. We use SDO/AIA 304 \AA~and 
171 \AA~data to study the filament evolution and its kinematics as it provides limb view for this event with less projection effect. 
SECCHI EUVI data (304 \AA,~195 \AA,~and 171 \AA) have been used to study the on-disk view of the filament evolution 
in the particular eruptive region. We also use LASCO C2 data to study the CME kinematics and the downflows of its 
core in the outer solar corona. 

Fig.~1 shows the full disk images of the Sun in the STEREO/EUVI 171 \AA~and SDO/AIA 304 \AA. The boxes in these images 
show the location of the filament. Observations from both the STEREO-B/SECCHI and SDO/AIA provide unique opportunity to 
study the filament eruption on the disk as well as on the limb respectively. The filament is located at the solar disk on the South-Western 
side as indicated by the box in the field of view of STEREO-B (Fig.~1, left image). On the same time, SDO/AIA observes the filament on 
the South-Eastern limb of the Sun (Fig.~1, right image). In the subsequent sections, we discuss the observations of the highly 
asymmetric filament eruption and the two-ribbon flare, as well as their association with the CME and the downflow of 
its core in the outer corona. Table~1 represents the time-line of the whole event started from the filament eruption, and end with the 
 downflows of the CME core in the outer corona.
\subsection{STEREO-B/SECCHI and SDO/AIA Observations of Asymmetric Filament Eruption and Two Ribbon Flare}
\label{sec2.1}
Fig.~2 shows the sequence of selected EUV 304 \AA~images ($T_{f}=0.07 MK$) of the filament evolution as observed by 
STEREO-B/SECCHI. A long dark filament has been clearly seen on the South-West hemisphere over the solar disk 
(cf., snapshot at 20:26:36 UT). The leading edge (north-west part) of the eruptive filament is indicated by the white arrows. 
Thereafter, the whole filament exhibits whipping-like asymmetric eruption. A two ribbon flare 
was observed underneath the southern part of the former filament position. The flare was un-classified in the GOES X-ray fluxes
because it occurred on the invisible side of the Sun. The two ribbons of the flare (indicated by R1 and R2 at the snapshot of 21:06:36 UT) are 
clearly visible. The flare occurred when the eruptive filament accelerated from slow speed to high speed (Fig.~7). For the 
detailed discussions of its physical scenario, we refer to the Section 3.1. The flare ribbons separate with each other along with the progress of
the filament eruption. The north-western leg of the filament remains anchored on the Sun, while the whole material in 
the filament channel first whipped and later erupted asymmetrically. Fig.~3 shows the sequence of the selected EUV 195 \AA~($T_{f}=1.4 MK$) images of the filament evolution observed by STEREO-B/SECCHI. The dark filament is also observed over the disk on the 
south-west of the solar disk (see the snapshot at 20:50:51 UT). In the coronal images, the fine and brightened coronal magnetic 
flux-tubes that cross the southern part of the filament, are clearly visible. The overlying magnetic field lines are clearly visible in this wavelength 
as indicated by the red arrows (cf., images at 21:00:51 and 21:05:52 UT in Fig.~3). These arches may be the part of a moderate active region 
lying on the eastern side of the filament (not shown here). The white arrows indicate the evolution of the leading edge (the
north-west segment) of the filament in the corona. The two ribbon flare, and asymmetric whipping like
eruption of the filament with its detached (southern) and anchored (northern) legs, are visible in 21:20:51 UT snapshot.

Fig.~4 and Fig.~5 respectively represent the sequence of the images in SDO/AIA 304 \AA~($T_{f}= 0.05 MK$) 
and 171\AA~($T_{f}= 0.6 MK$) channels showing the evolution of the 
filament eruption. The whole filament made-up of several threads is already visible on the south-east limb of the Sun on 2012 June 17. 
The white arrows indicate the leading edge of the eruptive filament. The observations show the asymmetric evolution and whipping 
of the filament (cf., filament.mpg). 
The whipping of the flux rope took place between $\approx$21:04 and $\approx$21:10 UT. During the whipping motion, most 
of the filament plasma channeled from the southern part to its northern part. The whipping and asymmetric eruption of the filament is 
indicated by the yellow arrow ( cf., 21:10 UT snapshot in Figs.~4 and 5). After the whipping motion, the filament  erupted 
asymmetrically higher in the corona towards the northern side of the Sun. The post flare loops in the southern side are visible during 
the decay phase of the flare (cf., 22:10:08 UT snapshot in Fig.~4 and 23:00 UT snapshot in Fig.~5). It is also interesting to note 
in the AIA 171 \AA~coronal image that the filament plasma shows some heating in the form of plasma brightening in 
the eruptive part. This may give a clue about the energy deposition and initiation of whipping-like asymmetric eruption through the magnetic 
reconnection in its southern part. This observational scenario matches well with the on-disk scenario of this filament eruption as observed 
by STEREO. The schematic representation of the asymmetric whipping like filament eruption and its comparison with SDO/AIA 
304 \AA~image at 21:06:08 UT is shown in Fig.~6. The upper panel of Fig.~6 shows the SDO/AIA 
304 \AA~image of the filament eruption at 21:06:08 UT. The arrows indicate the two legs of the filament. The filament image is rotated 
by 90 degree clockwise from its real position to compare with the schematic diagram. The bottom panel of Fig.~6 shows the 
schematic of the whipping like asymmetric filament eruption. Red line indicates the polarity inversion line.

Fig.~7 represents the height-time plot of the filament using the SDO/AIA 304 \AA~image sequence. The filament shows a slow rise 
followed by a fast rising phase. The speed of the slow rise is around 3 km s$^{-1}$, while the fast rising phase has a speed of 
about $\approx$205 km s$^{-1}$. The filament rises slowly from 20:30 UT to 20:56 UT, and 
thereafter exhibits a fast rising phase from 21:00 UT to 21:14 UT. The whipping-like asymmetric eruption of the filament took place around 
its transition from the slow rise to the fast rise phase (Sterling \& Moore 2005). After the whipping motion, the filament erupted 
with a high speed. The speed has been calculated from the linear fit to the height-time data. 
When we examine the composite image (Fig.~8) of the SoHO/LASCO white light image on 2012 June 17 at 23:24 UT 
and the SDO/AIA 304 \AA~image at 21:15 UT (inside), it shows occurrence of high degree of asymmetric filament 
eruption that later constitutes CME core. The detailed outer coronal dynamics is described in the next subsection.
\subsection{LASCO C2 Observations of the CME and Downflow of Its Core in the Outer Corona}
\label{sec2.2}

The LASCO C2 coronagraph observed the CME associated with the asymmetrical eruptive filament on 2012 June 17-18 
during 21:36-04:48 UT. It was a slow CME and appears from the east side of the coronagraph occulting disk. Fig.~9 shows sequence 
of the difference images, showing the CME and its core eruption. Thereafter, the CME core falls down, which initiates the outer 
coronal downflows (cf., cme-downflow.avi). The CME first appears in the LASCO C2 field of view at $\approx$21:36 UT on 2012 June 17. The red arrows in the panels (a)-(c) of Fig.~9, indicate the outward motion of the CME's leading edge. The core of the CME, which is the filament flux 
rope, was initially observed in the LASCO C2 field of view at $\approx$22:12 UT. The position angle and the width of the CME were 
$\approx$90$^{o}$ and $\approx$125$^{o}$ respectively. 

Fig.~10 shows the height-time plot of the CME leading edge, the CME core (Fig.~10a), and the derived velocity and acceleration 
profiles of the core with time (Fig.~10b), as well as the same profiles with distance (Fig.~10c). The solid black line 
shows the linear fit of the form $y(t)=a+bt$ to the measured height of the CME leading edge w.r.t. time.
From the linear fit, the speed $(dy(t)/dt=b)$ of the CME leading edge is estimated as $\approx$540 km s$^{-1}$. The 
dotted blue line shows the height of the CME core w.r.t. time. From the linear fit to the upward motion of CME core, we 
obtain its average velocity as $\approx$126 km s$^{-1}$. Using the same method, we estimate the average velocity of $\approx$ 
56 km s$^{-1}$ during the core downflow. In order to determine the velocity and acceleration at each point of the parabolic 
path of the CME core, we perform the third order fit. The blue dotted line over these points shows the third order fit of the form 
$y(t)=a+bt+ct^{2}+dt^{3}$. The velocity and acceleration profiles are derived by differentiating the above equation at each point 
along the height-time profile (Fig.~10b). It is clear from Fig.~10b that the velocity of the CME core attains zero value 
at $\approx$00:48 UT on 2012 June 18 when it reaches the maximum height. 
The CME core shows continuous deceleration. The deceleration rate changes from the initial value of -48 m s$^{-2}$ to -19 m s$^{-2}$ 
at the maximum height. Thereafter, it starts downflowing with acceleration rate from $\approx$-21 m s$^{-2}$ at 00:48 UT 
to $\approx$14 m s$^{-2}$ at 03:48 UT (cf., velocity and acceleration profiles in Fig.~10b). We also plot the velocity and 
acceleration profiles of the CME core w.r.t. radial distance. The initial velocity at $\approx$2.54 $R_{\sun}$ is $\approx$312 km s$^{-1}$, 
which continuously decreases with height and becomes zero at $\approx$4.33 $R_{\sun}$ (cf., Fig.~10c). After the start of 
downflow from $\approx$4.33 $R_{\sun}$, its initial speed is around $\approx$10 km s$^{-1}$ that later reaches up 
to $\approx$30 km s$^{-1}$ at 3.56 $R_{\sun}$. The initial deceleration was $\approx$48 m s$^{-2}$ at the height 
of $\approx$2.54 $R_{\sun}$. At 4.33 $R_{\sun}$, the deceleration was 21 m s$^{-2}$, while during the fall of the core 
from this height, it reaches to zero and then tends towards positive values.  
The value of deceleration/acceleration during the motion of the core was not matching to the local gravitational acceleration 
$GM_{\sun}/r^{2}\sim68 (2R_{\sun}/r)^{2} m/s^{2}$ (Wang \& Sheeley 2002) (cf. Fig.~10c). This evidently shows that the gravity is not the dominant factor for the observed downward motion. We also observe the interaction of the CME with the coronal ray in the northward 
direction of the outward moving core. The red curve in the Fig.~10a shows the coronal ray deflection from its mean position at a height of  
$\approx$3.1 $R_{\sun}$. The deflection started around 22:12 UT on 2012 June 17 soon after the appearance of CME core in the 
LASCO field of view and continued with the upward motion of the CME. The coronal ray is deflected maximum of 3.2$\times 10^{5}$ km 
at $\approx$22:36 UT in the north and then returns back to its mean position. 
\section{Physical Scenario of the Observed Dynamics}
\label{sec3}
The observations of the asymmetric filament eruption, formation of the two ribbon flare, and their relationship with the CME and the outer coronal downflows are discussed in detail in the previous section. In the present section, we describe the possible physical scenarios and theoretical interpretations for the observed coronal dynamics.
\subsection{Asymmetric Filament Eruption and Two Ribbon Flare}
\label{sec3.1}
The asymmetric flux-rope eruptions are studied by various authors (Tripathi et al. 2006; Liu et al. 2009, and reference therein). We observed the 
whipping-like asymmetric filament eruption. The whipping like asymmetric filament eruption is characterized when active leg whips 
upward, and hard X-ray sources shift toward the end of the anchored leg (Liu et al. 2009). We do not have the X-ray observations in our case. 
However, we could not find any evidence of the transport of flare ribbon brightening (low energy counterparts) towards the encored leg of 
the filament. This is some what unique situation differing from the observations of Liu et al. (2009). In the present observational base-line, it is 
clear that the northern part of the filament has comparatively homogeneous  overlying corona, while its southern part experienced the 
crossing overlying thin and brightened flux-tubes. In the south-east direction of it, there was a moderate active region whose arches were 
crossing above the southward part of the filament and made it initially suppressed. During the whipping motion of the southern part of 
filament these overlaying magnetic flux-tubes probably allow only the formation of a compact two ribbon flare beneath the eruptive part 
of the filament. This flare brightness does not spread much along the neutral line as in the case of large flares (Benz 2008).

The filament erupts asymmetrically and the two ribbon flare occurs beneath of it. The flare may be triggered due to reconnection 
in the vertical current sheet between surrounding coronal fields in the wake of the eruptive filament (Carmichael 1964; Sturrock
1966; Hirayama 1974; Kopp \& Pneuman 1976; Shibata 1999, reference cited therein). As the reconnection progresses as per those canonical models, 
formation of post-flare loops, propagation of the brightening along the ribbons as well as their separation are evident in the lower part 
of the solar atmosphere.

\subsection{The coronal Downflows due to the Flux-rope Failed Eruption.}
\label{sec3.2}
The eruptive filament further leads to a CME whose core is constituted by it. The degree of asymmetry in this eruption is large. It is clear from the Fig.~8 that inspite of radial motion, the filament expands in the outer corona (CME core) almost at an angle of $\approx50^{\circ}$ and constitutes the outer coronal dynamics in form of downflow. The core later on shows the downflow of its plasma. The physical reason in the present observational base-line may be two-fold, e.g., (i) The self-consistent equilibrium of the flux-rope below a critical height, during
its subsequent eruption in form of whipping motion and formation of the CME's core (Filippov et al. 2001, 2002). 
(ii) The interaction of the moving CME core with the ambient medium that drags 
out its momentum (Wang \& Sheeley 2002). Before the start of downflow, CME core causes the deflection of a coronal ray in the northward direction 
in the LASCO C2 field-of-view (cf. Fig.~9). The deflection measurement has been carried out at a height of around $\approx$3.1 
$R_{\sun}$. The ray started deflecting around 22:12 UT and reaches up to a maximum displacement of $3.2\times10^{5}$ km at 
$\approx$22:36 UT on 2012 June 17 (cf., Fig.~10a). This was the period when CME core was trying to move ahead with the average 
velocity of $\approx$126 km s$^{-1}$ (cf., Fig.~10b). The coronal ray deflection and its re-storing may be most likely caused by 
the expansion of the CME-core magnetic field and its interaction with the radial field of the coronal ray (Filippov \& Srivastava 2010). However, during the 
coronal ray restoring motion, the downflow of CME core starts when the coronal ray moved almost 50\% towards its equilibrium position 
on $\approx$ 01:00 at 18 June 2012 (cf. Figs.~9 and 10a). This was the time when CME core starts down flowing with an average speed of around 
56 km s$^{-1}$. Both the southward opposite motion of coronal ray (in projection), and coronal downflow, were progressed simultaneously 
later on, and finally ended at $\approx$ 03:48 UT on 2012 June 18. This interaction and co-temporal dynamics of the coronal ray and coronal 
downflow is, to the best of our knowledge, observed for the first time in the presented observations. However, the one-to-one relation 
between the coronal ray restoring and downflow is not established here as the downflow starts after two hour from the 
initiation of restoring motion of the deflected coronal ray towards its equilibrium position. Therefore coronal ray deflection may be only the consequence of the upward moving CME, however, it does not start the downflow of CME core. 

The most likely cause may be the self-consistent evolution of the flux-rope itself, 
which forms the CME core. We compare the observed evolution of the 
eruptive filament and CME with the model developed by Filippov et al. (2001, 2002) for the non-radial 
flux-rope motion previously. It was applied to the event on 1997 December 14 whose initial phase was very similar to the eruption 
studied in the present paper. In the December 14 event, the filament erupted from middle latitude in the southern hemisphere, while the CME 
was observed close to the equatorial plane. In the present case, the CME core propagates even at some angle to the north of the equatorial plane 
(cf. Fig.~8). In both cases, the filaments passed significant distances along the latitude direction. However on 1997 December 14, the true fast 
CME was observed, while in the present case, on 2012 June 17-18, the CME core is stopped at some height and then fall down.

In the axially-symmetric model of Filippov et al. (2001, 2002), the flux-rope was represented by a thin current-carrying plasma ring (torus) with a total electric current $I$ located above the photosphere along a heliographic parallel. A similar model for a ring located in the equatorial plane was analyzed by Lin et al. (1998). After integrating over the torus volume, we can obtain the equation of motion of the torus as a whole, as well as an equation describing variation of its inner radius. In cylindrical coordinates ($\rho$, $\varphi$, $z$) with their origin at the solar center and the z axis directed along the rotational axis, the equations of motion for a toroidal segment with unit length take the form

\begin{equation}
m\frac{d^2\rho}{dt^2} = \frac{I}{c} (B_z^{(ex)}+B_z^{ (m)}+B_z^{(I)}) - mgR^2 \frac{\rho}{(\rho^2+z^2)^{3/2}} - kv_p,
\end{equation}
      
\begin{equation}
m\frac{d^2z}{dt^2} = -\frac{I}{c} (B_\rho^{(ex)}+B_p^{(m)}) - mgR^2 \frac{z}{(\rho^2+z^2)^{3/2}} - kv_z,
\end{equation}
      
where $m$ is the mass of the filament per unit length, $B^{(ex)}$ is the magnetic field produced by sources located beneath the photosphere and by currents in the solar-wind region, $B^{(m)}$ is the field produced by inductive currents in the photosphere that prevents the penetration of the coronal-current field into the Sun, $B^{(I)}$ is the field produced by the current flowing along the ring axis, $g$ is the free-fall acceleration at the photospheric level, $k$ is the dissipation coefficient, and $v$ is the velocity.

We used the same parameters of the model as in Filippov et al. (2002) except the mass of the filament per unit length. It was chosen as $10^{5}~g~cm^{-1}$, a more typical value for an average filament, because we do not need to fit the result to the final velocity of $500~km~s^{-1}$. The global coronal field $B^{(ex)}$ is represented by two spherical harmonics, dipole and octupole. These harmonics dominate in the global field in the epoch of solar activity minimum. Our event does not occurr at minimum, however, the level of activity in June 2012 was not very high as the monthly averaged sunspot number was about 60, thus the global magnetic field configuration was not typical for the maximum. The harmonic coefficients are chosen in such a way that the magnetic field strength is $B\approx1~G$ in the corona near the filament, and there is a null line at the height of $0.2~R_{\sun}$ above the equator. In this field, a flux rope with the current of $\approx~10^{11}~A$ will be in equilibrium at the height of $\approx30~Mm$. The equilibrium is defined mainly by the components of the Lorenz force in Equations (1) and (2). The gravity force is about two orders of magnitude weaker than the electromagnetic forces acting on the filament. It does not appreciably affect the equilibrium position in the lower corona, however the filament mass is significant for the kinematics of eruption. 

Each value of the current corresponds to two equilibrium points in this curve - one above and one below some critical point whose current $I_c$  corresponds to a single solution of Equations (1) and (2) with zero left-hand terms. If the current exceeds the critical value, equilibrium cannot be achieved in the corona. The left panels of Fig.~11 shows the evolution of the flux rope without dissipation after the loss of equilibrium due to the increase of the electric current strength on a small value (about 2\%) over the critical value. We see that the true flux rope eruption started in the southern hemisphere and propagating into the northern hemisphere. 

It was pointed out by Filippov et al. (2001) that there is a stable equilibrium position higher in the corona for the erupting flux rope, but the gained kinetic energy of the flux rope does not allow it to stop in the higher equilibrium position. However, if we add some dissipation $-kv$ to the right hand side of the equations of motion, as it was done by Filippov et al. (2001) for testing of the initial equilibrium stability, the flux rope can loose its momentum and stop at some height (the right panels of Fig.~11). We choose the value of $k$ as $2 \times 10^{2}~g~sec^{-1}$ in order to stop the ascending motion of the flux rope at the height of 4 $R_{\sun}$. At the speed of $300~km ~s^{-1}$, it creates the drag force $kv = 6\times10^{9}$ dyne per unit length or about an order of magnitude stronger than the gravitational force $fg = 7\times10^{8}$ dyne at the height of 2 $R_{\sun}$. Although we did not apply additional efforts to fit to the observational data, we can see that the kinematics shown in the right panels of Fig.~11 is very similar to the observed kinematics presented in Fig.~10. We compare the segments of curves between the dashed vertical lines in Fig.~11d with the dotted blue line fitting in Fig. 10a and the segments of curves between the dashed vertical lines in Fig.~11e with the black dotted line connecting the data points in Fig. 10b. They show 6-hours interval of the CME core evolution. Evidently, the curves are very similar and the quantities are close to each other. Comparison of Fig.~8 with the calculated trajectory in Fig.~11f also confirms the conclusion that the model rather adequately describes the observed event. However, the nature of the dissipation or the drag force is unclear for us. Possibly it is the aerodynamic drag or includes some additional dissipative mechanisms.
\section{Discussion and Conclusions}\label{sec5}
\label{sec4}
We have presented the multi-wavelength analysis and the relationship of whipping-like asymmetric filament eruption, the associated CME, and the outer coronal downflows in the form of CME core on 2012 June 17-18. The summary of the obtained results of this study is as follows:\\
\\
1. The observation shows a whipping like highly asymmetric filament eruption with active (south-eastern) leg and the anchored (north-western) leg.\\
2. During the eruption, a two ribbon flare occurred underneath the eastern part of the filament. This supports well the standard flare model (CSHKP) of reconnection.\\ 
3. The deceleration profile of the CME core shows that the gravity is not only the force responsible for the downflow. The downflow of the CME core has been observed which may be due to the self consistent evolution of the flux-rope in the coronal magnetic field.\\
4. Coronal ray deflection occurs during the upward motion of the CME, however, it does not reveal exact correlation with the coronal downflows.\\ 

Recently Liu et al. (2009) presented observation of two types of asymmetric filament eruptions, i.e., whipping and zipping like configurations. In the whipping like eruptions, active leg whips upward, and the observed hard X-ray source locations shift towards the anchored leg. While in the zipping like asymmetric eruption, active leg moves along the neutral line and the hard X-ray sources move away from the anchored leg. 
In our present observations,  the filament eruption is whipping-type highly asymmetric eruption, however, without any significant propagation of flare brightening  towards its anchored leg. Whipped filament also rises from nearby its southern activated footpoint, and energy deposition in the form of  flare ribbons occurred only beneath of it.  The asymmetric eruption of the filament later produced a slow CME that deflects the coronal ray in the outer corona. Thereafter, the coronal downflow is observed from $\approx$4.33 $R_{\sun}$ towards the Sun. Previously, Wang \& Sheeley (2002), observed the CME core fallback during 1998-2001. They found that the fallback events occur in impulsive but relatively slow CMEs. The fall down of core material may be due to its interaction with the background plasma which removes the momentum of the CME core. McKenzie \& Hudson (1999)
 observed a downward flow above the coronal arcades during the decay phase of solar flare on 1999 January 20, and interpreted it as cross sections of evacuated flux-tubes resulting form the intermittent reconnection following the associated CME. However, in the present case, we have a moderate CME beyond 2 $R_{\sun}$ with its core, formed by highly asymmetric filament eruption that may start downflowing due to its self-consistent evolution in the coronal magnetic field as shown in our model calculations (Fig.~11). 

In conclusion, our present multiwavelength study provides a unique relationship between the asymmetric filament eruption, flare as per standard model, associated CME, and the initiation of coronal downflows around $\approx$4.33 $R_{\sun}$. Such studies may have important implications on the  physical conditions of the outer corona and its coupling through various types of transients and plasma dynamics occurred in the lower solar  atmosphere. New observations should be performed using future high-resolution observations from space and ground to shed new light on such significant phenomena and their relationship in the solar atmosphere.
\acknowledgments
We thank anonymous referee for his/her valuable comments and suggestions. The authors thank Indo-Russian (DST-RFBR) Project on "Study of Dynamical Events in the Solar Atmosphere during Maximum of Solar Cycle 24 (INT/RFBR/P-117 RFBR 12-02-00008 and 12-02-92692)" for its support to this study during our bilateral collaboration. NCJ thanks Aryabhatta Research Institute of Observational Sciences (ARIES), Nainital for providing Post Doctoral Grant. BF was supported in part also by the Program \# 22 of the Russian Academy of Sciences. We also acknowledge partly the support of IUSSTF/JC-Solar Eruptive Phenomena/99-2010/2011-2012 project. AKS thanks Shobhna Srivastava for the patient encouragements for his research works.


\clearpage
\begin{figure}
\centerline{
	    \includegraphics[width=1.1\textwidth,clip=]{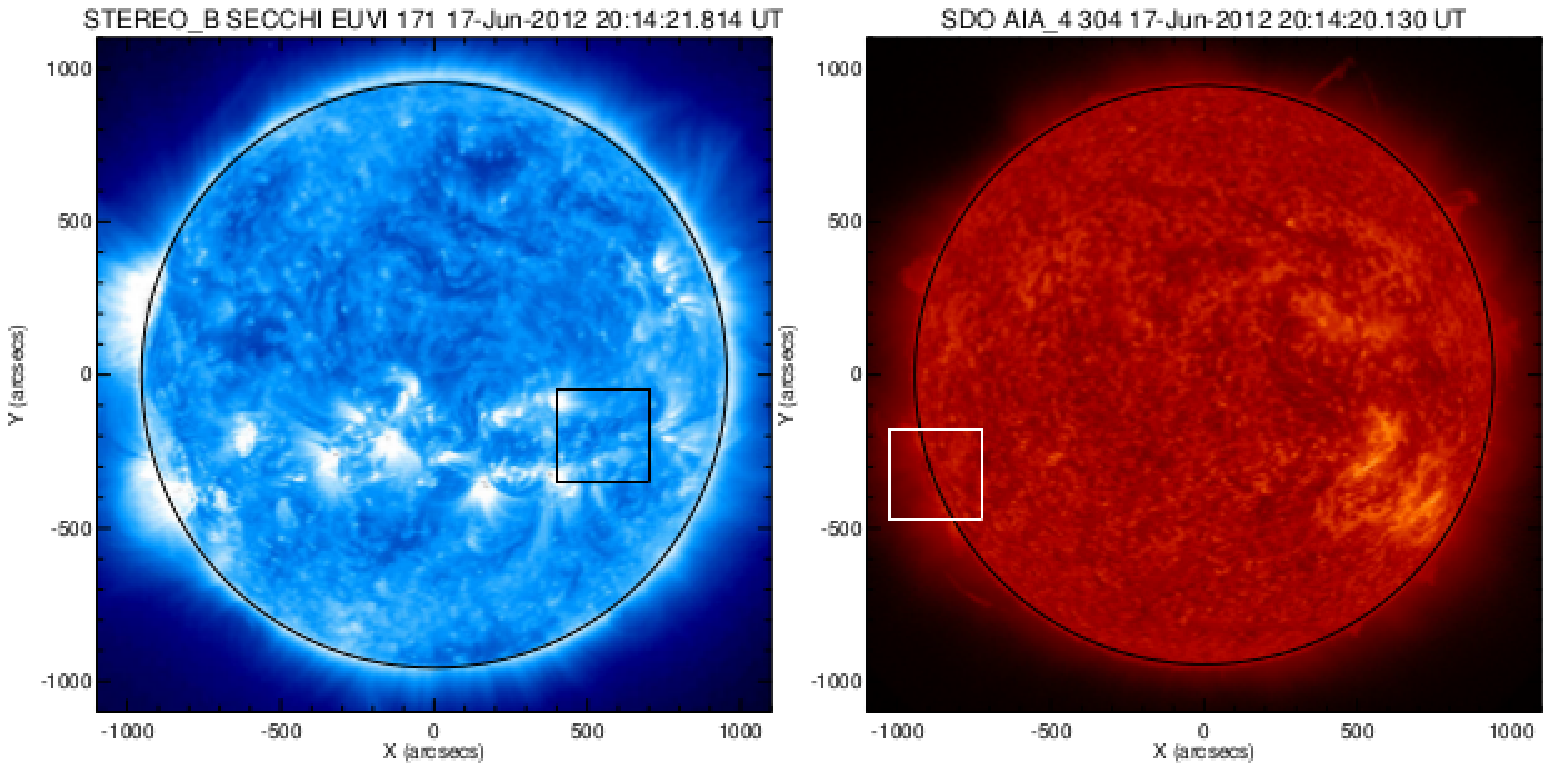}
	   }
\vspace*{-0.5cm}
\caption{STEREO-B/SECCHI 171 \AA~(left) and SDO/AIA 304 \AA~(right) images showing the full disk of the Sun on 2012 June 17. 
The boxes show the location of the filament from two different angles.}
\label{fig1}
\end{figure}
\clearpage
\begin{figure} 
\centerline{
	    \hspace*{0.0\textwidth}
            \includegraphics[width=1\textwidth,clip=]{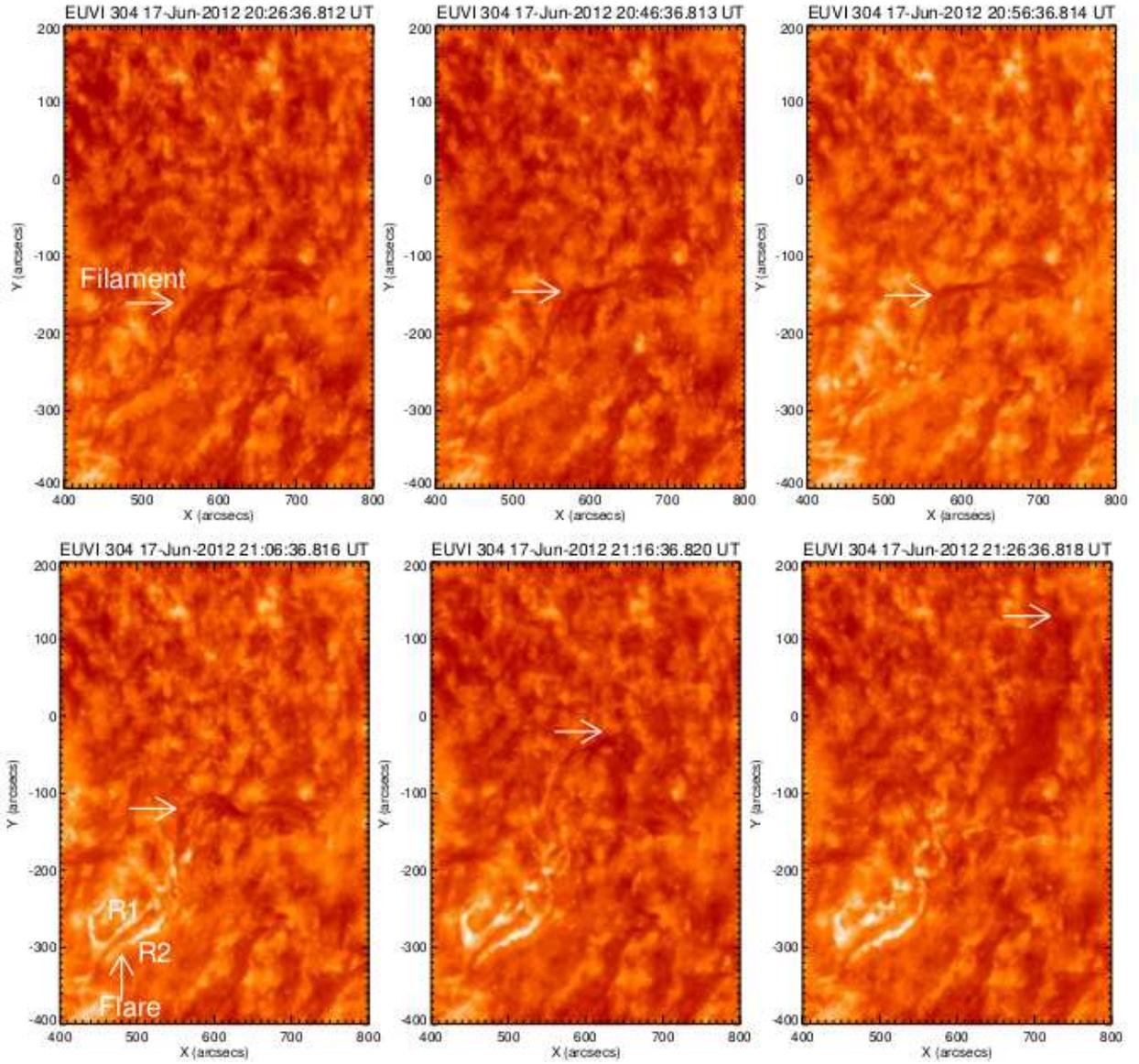}
           }
\vspace*{-0.5cm}
\caption{STEREO B/SECCHI 304 \AA~images showing the activation and eruption of the filament and the triggering of a two ribbon flare near the South-Eastern foot point. The two ribbons of the flare are represented by R1 and R2.} 
\label{fig2}
\end{figure}
\clearpage
\begin{figure} 
\vspace*{-1cm}
\centerline{
	    \hspace*{0.0\textwidth}
            \includegraphics[width=1\textwidth,clip=]{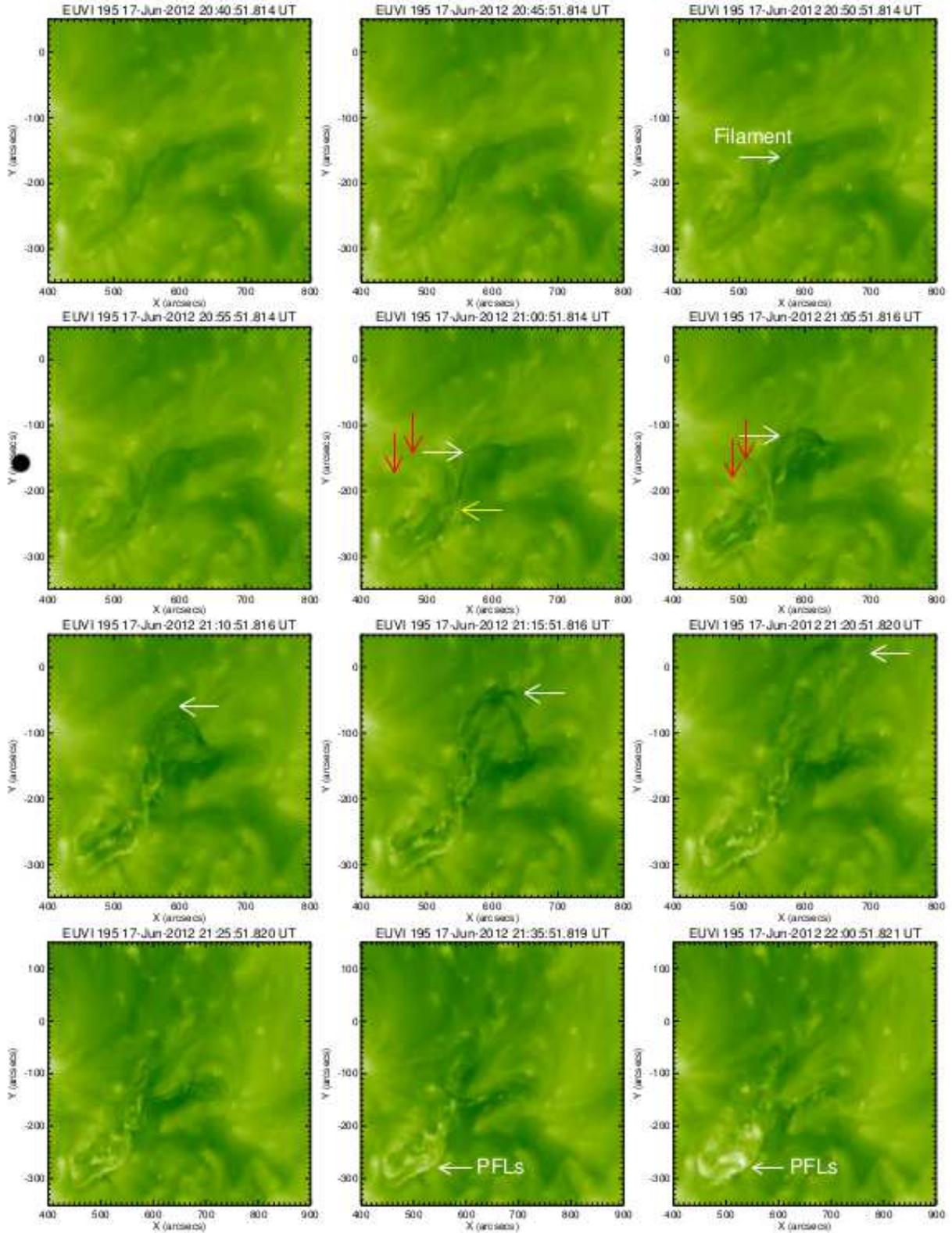}
	   }
\vspace*{-0.5cm}
\caption{STEREO-B/SECCHI 195 \AA~images showing the activation of the filament. The white arrows represent the eruption of the filament.
Downward red arrows indicate the overlaying magnetic field arches above the southern part of the filament. The yellow arrow indicates the interaction of filament with the overlaying magnetic field. Post flare loops (PFLs) are presented in the bottom panel.}
\label{fig3}
\end{figure}
\clearpage
\begin{figure} 
\vspace*{-1cm}
\centerline{
	    \hspace*{0.0\textwidth}
            \includegraphics[width=1\textwidth,clip=]{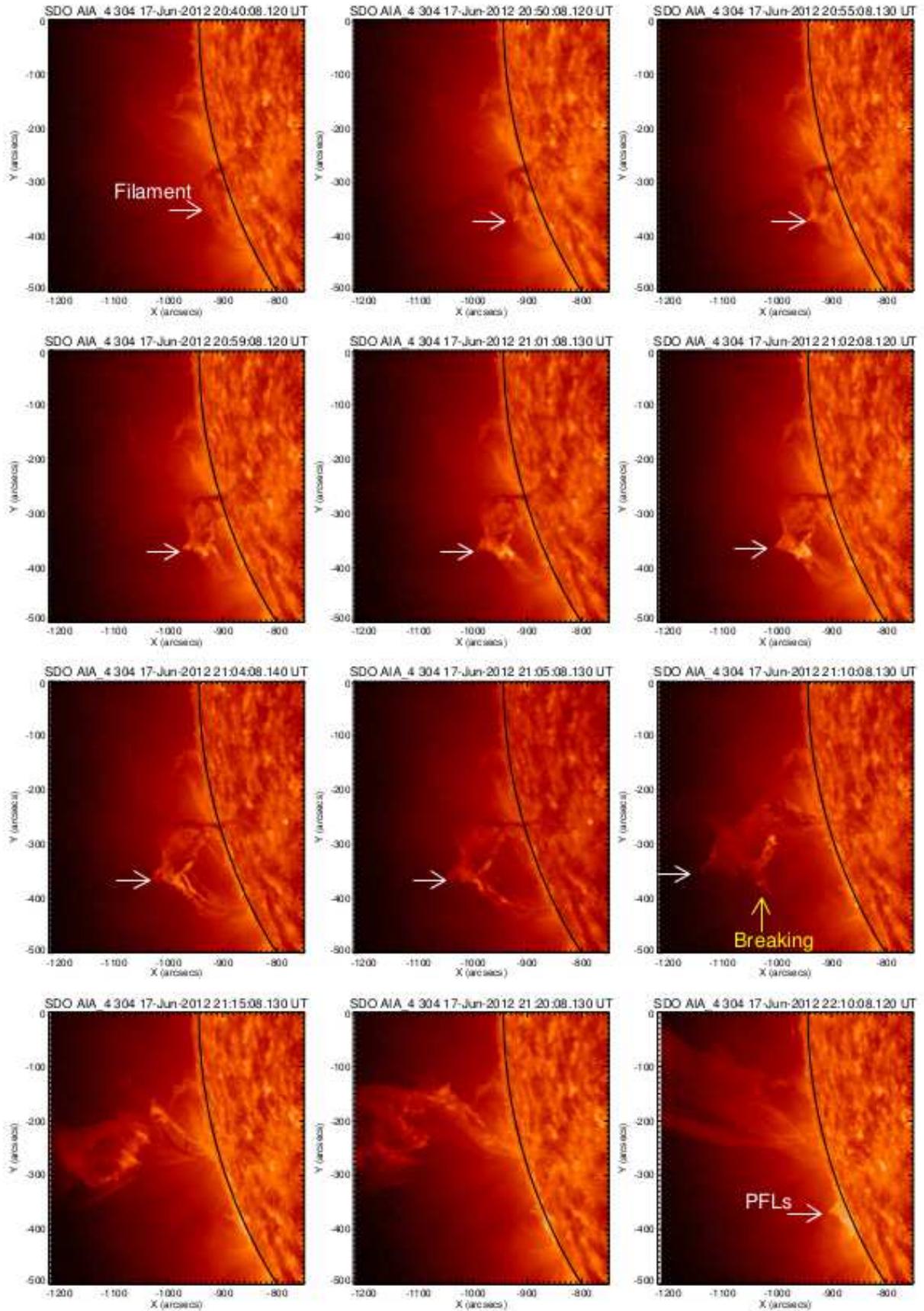}
	   }
\vspace*{-0.5cm}
\caption{SDO/AIA 304 \AA~images showing the evolution and eruption of the filament. White arrows indicate the erupting leading edge of the filament. The bottom right corner image shows the post flare loops (PFLs).}
\label{fig4}
\end{figure}
\clearpage
\begin{figure} 
\vspace*{-1cm}
\centerline{
	    \hspace*{0.0\textwidth}
            \includegraphics[width=1\textwidth,clip=]{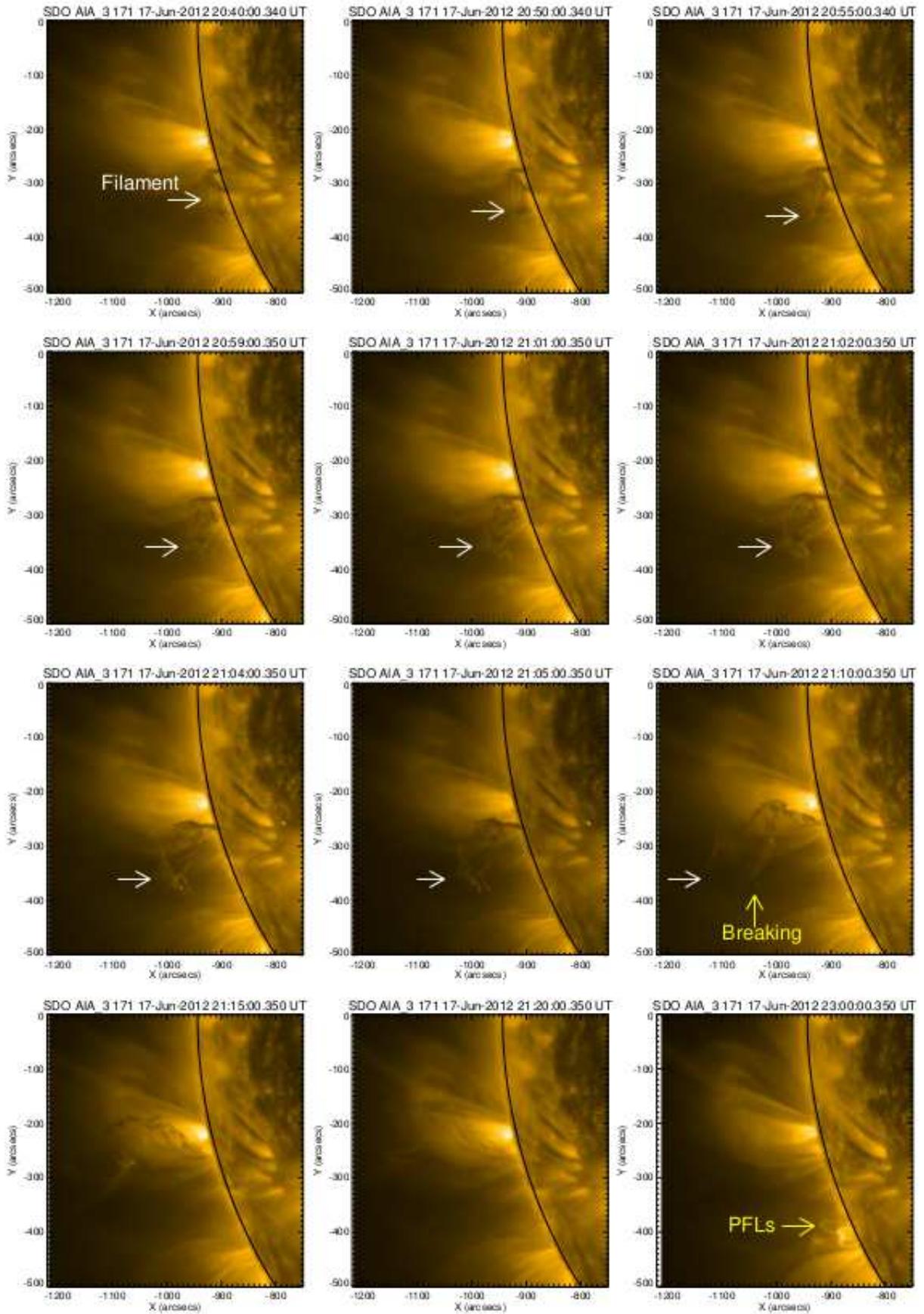}
	   }
\vspace*{-0.5cm}
\caption{SDO/AIA 171 \AA~images showing the evolution and eruption of the filament. White arrows represent the evolution of the filament 
leading edge. Post flare loops (PFLs) on the southern part of the filament is clearly visible in the bottom right image.}
\label{fig6}
\end{figure}
\clearpage
\begin{figure}
\centerline{
	    \hspace*{0.0\textwidth}
            \includegraphics[width=0.7\textwidth,clip=,angle=0]{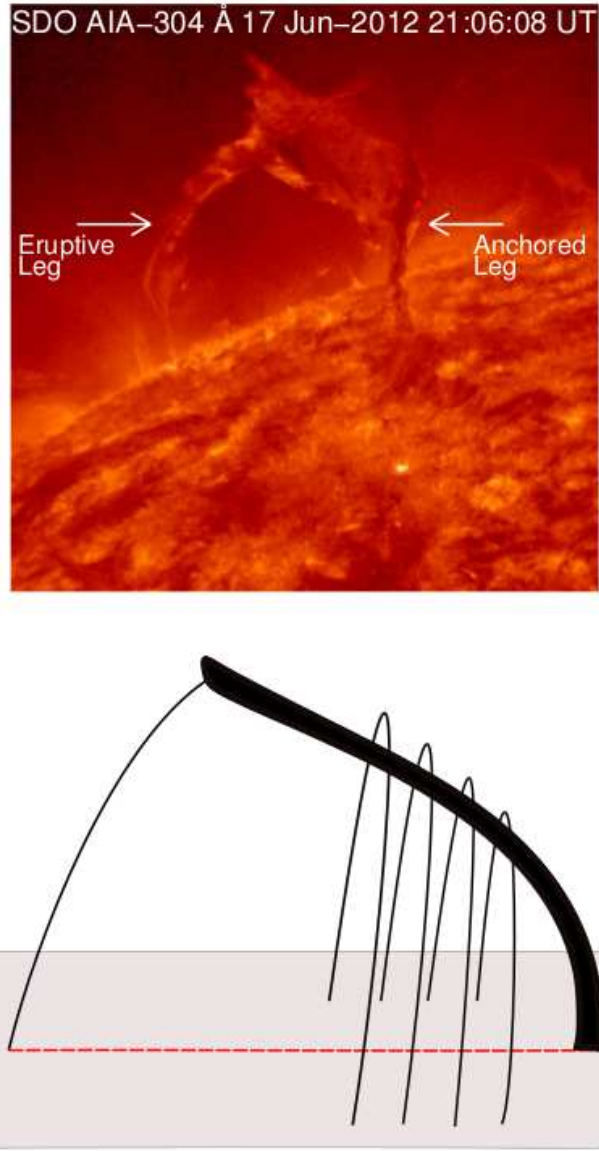}
	    }
\caption{Upper panel: SDO/AIA 304 \AA~image of the filament eruption at 21:06:08 UT. The arrows indicate the two legs of the filament. The filament image is rotated by 90 degree clockwise from its real position just to compare with the schematic diagram. Bottom panel: Schematic diagram of the whipping like asymmetric filament eruption. Red line indicates the polarity inversion line.}
\label{fig7}
\end{figure}
\clearpage
\begin{figure}
\epsscale{.8}
\plotone{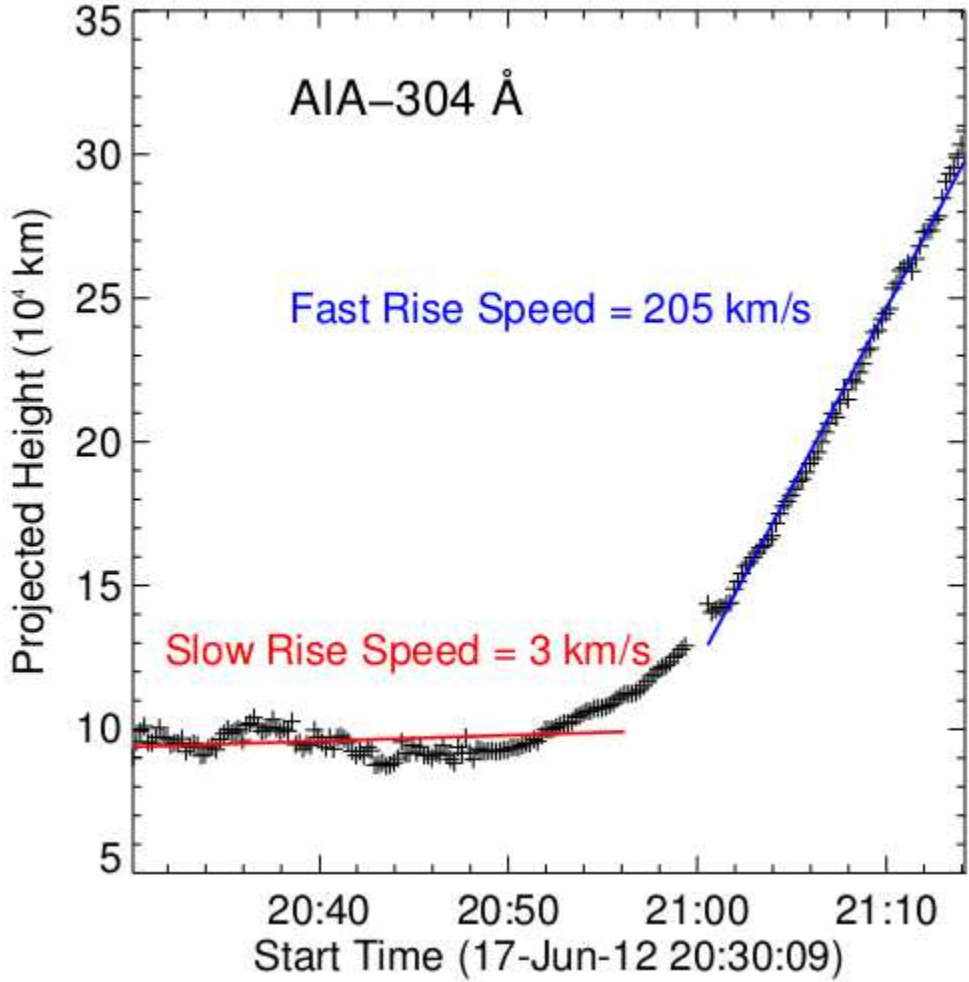}
\caption{Height-time plot of the filament leading edge (indicated by the white arrows in the Fig.~4) measured using SDO/AIA 304 \AA~time-series.}
\label{fig5}
\end{figure}
\clearpage
\begin{figure}
\epsscale{0.8}
\hspace*{-2cm}
\plotone{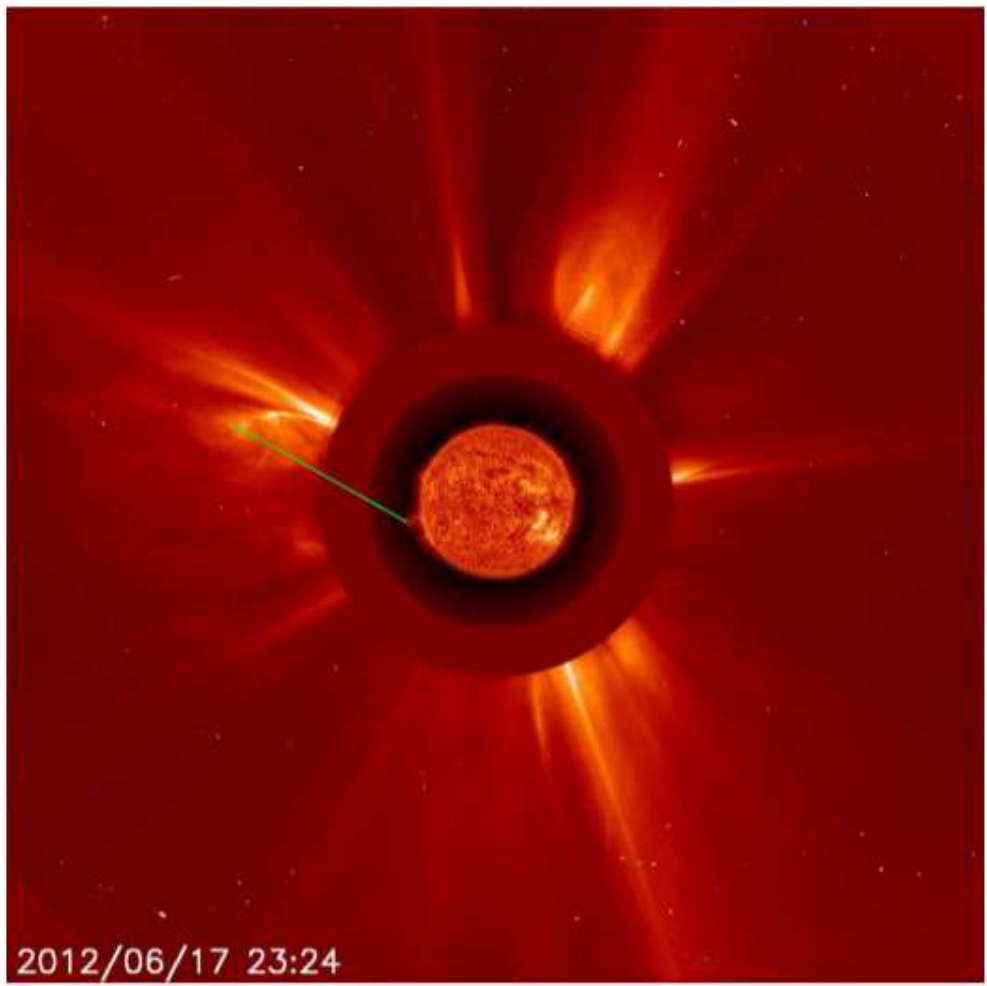}
\caption{Composite image of the white-light SOHO/LASCO-C2 image on 2012 June 17 at 23:24 UT and SDO/AIA 304 \AA~image at 21:15 UT (inside). Green line shows the approximate trajectory of the eruptive prominence.}
\label{fig10}
\end{figure}
\clearpage
\begin{figure} 
\centerline{\hspace*{0.0\textwidth}
            \includegraphics[width=0.9\textwidth,clip=]{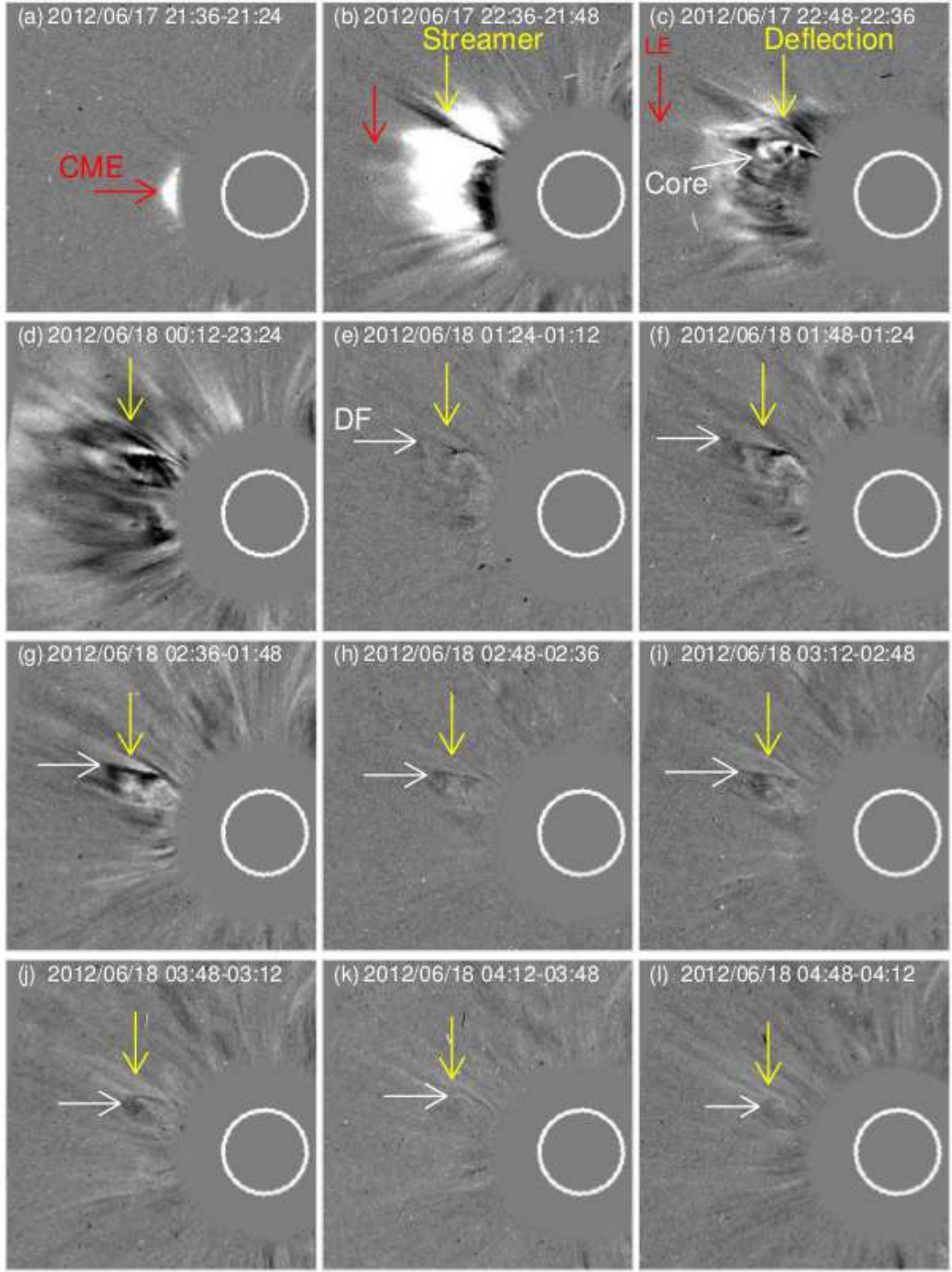}
	   }	    
\vspace*{0.5cm}
\caption{Time sequence of LASCO C2 images showing the CME eruption (a-d) and falling down of the CME core material (e-l) 
on 2012 June 17-18. Red arrows indicate the evolution of the CME leading edge (LE). White arrows indicate the down flow (DF) of the CME core. Yellow arrows indicate the coronal ray deflection. These difference images have been taken from SoHO LASCO CME CATALOG $(http://cdaw.gsfc.nasa.gov/CME_list/)$.}
\label{fig8}
\end{figure}
\clearpage
\begin{figure}
\epsscale{0.9}
\plotone{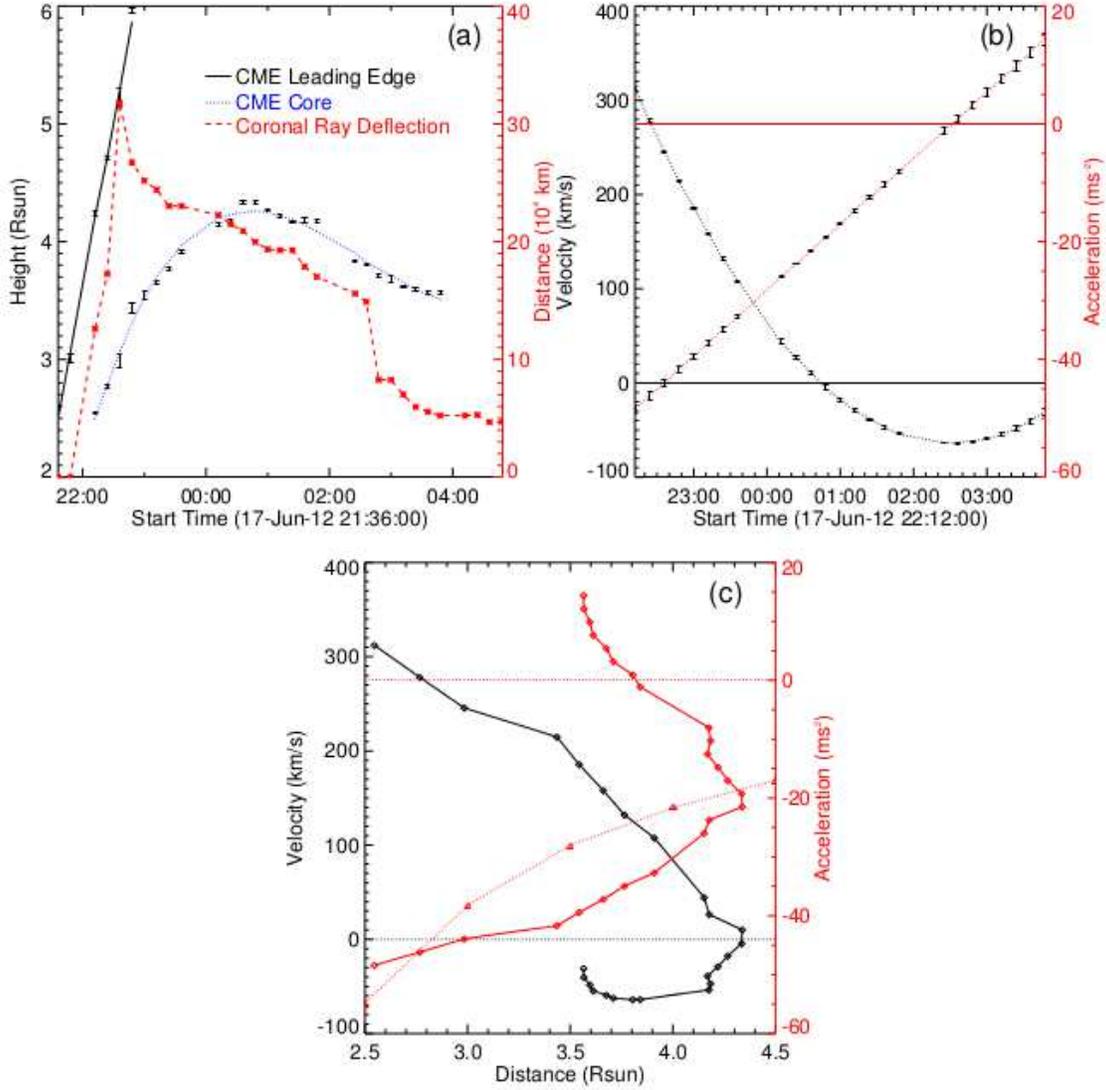}
\caption{(a) Height-time plot of the CME core and leading edge. The solid black line represents the linear fit to the CME leading edge data. The dotted blue curve in this figure represents the third order fit to the core height-time data. (b) Speed-time and acceleration-time plots of the CME core. Error bars in the panel (a-c) are the standard deviation computed from the three repeated measurements. (c) Distance-velocity and distance-acceleration plots  of the CME core. The velocity and acceleration are represented by the black and red colors respectively. The red dotted curve with triangle symbols in the panel (c) corresponds to the local gravitational acceleration $-GM_{\sun}/r^{2}$.}
\label{fig9}
\end{figure}
\clearpage
\begin{figure}
\epsscale{1}
\hspace*{-2cm}
\plotone{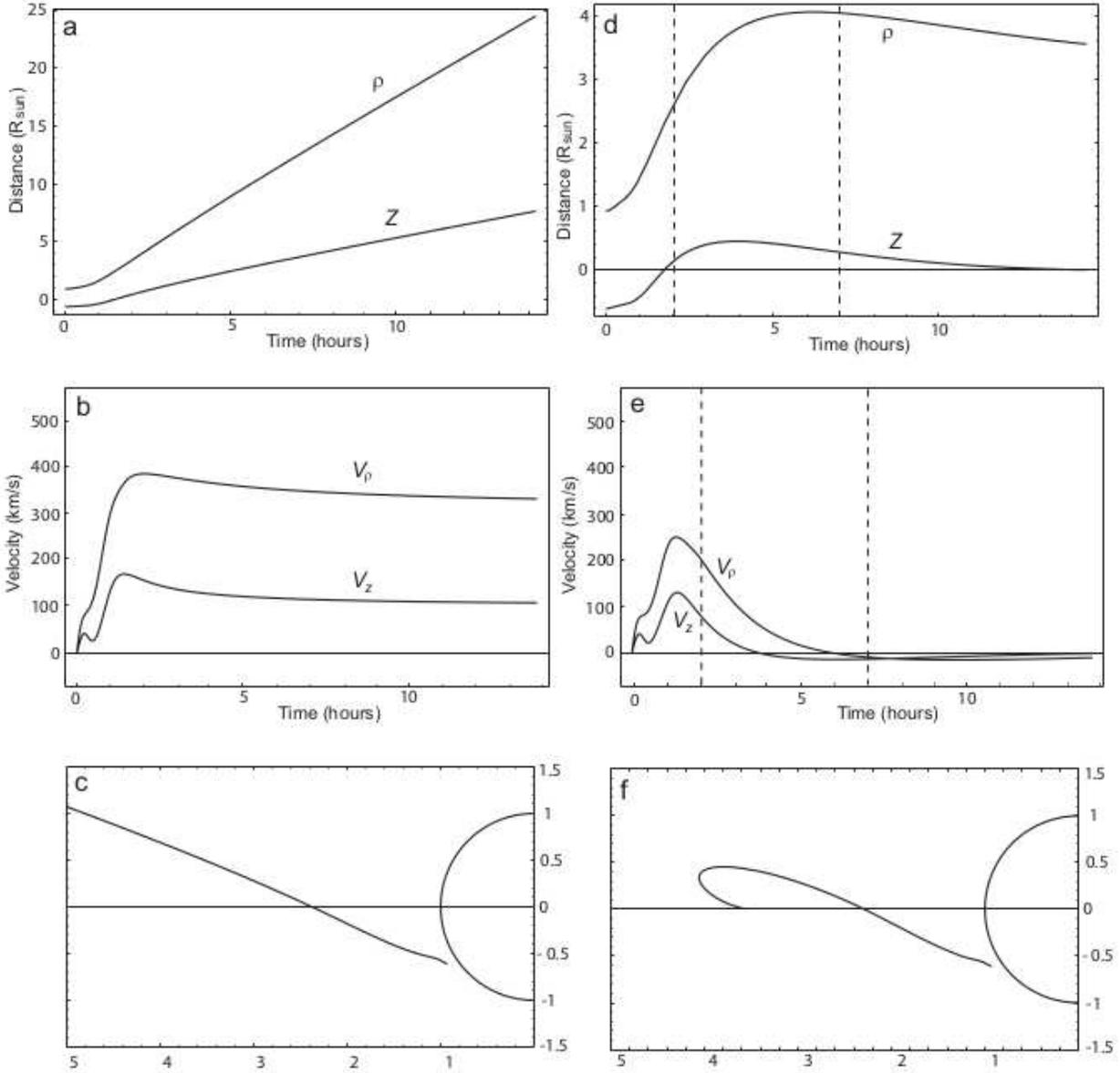}
\caption{The dependence of the flux-rope coordinates, z (top row) and the speed (middle row) on time and the trajectory of the erupting flux-rope motion (bottom row). Left panels correspond to the absence of dissipation, right panels describe the evolution with dissipation.}
\label{fig10}
\end{figure}
%
\clearpage
\begin{table}
\caption{Time-line of the Event.}
\label{table1}
\begin{tabular}{lll}     
\hline
S.No. & Date (Time) & Observations\\
\hline
1. & 17-06-2012 (before 20:30 UT) & Filament is observed on the Sun.\\
2. & 17-06-2012 ($\approx$20:30$-$$\approx$20:56 UT)  & Slow rise of the filament with the speed around 3 km s$^{-1}$.\\
3. & 17-06-2012 ($\approx$21:00$-$$\approx$21:14 UT) & Fast rise of the filament with the speed around 205 km s$^{-1}$.\\
4. & 17-06-2012 ($\approx$21:36 UT) & First appearance of the CME in the LASCO C2 field of view.\\
5. & 17-06-2012 ($\approx$22:12 UT) & First appearance of the CME core in the LASCO C2.\\
    &  & Start time of the coronal ray deflection.\\
6. & 17-06-2012 ($\approx$22:36 UT) & Maximum deflection time of the coronal ray.\\
7. &  17-06-2012 (after $\approx$22:36 UT) & Returning motion time of the coronal ray.\\
8. & 18-06-2012 (after $\approx$00:48 UT) & Start time of the downflow of the CME core.\\
\hline
\end{tabular}
\end{table}
\end {document}